\documentclass[journal,twoside]{IEEEtran}

\pdfoutput=1

\usepackage[dvipdfmx]{graphicx} 
\usepackage{bmpsize}

\usepackage{lipsum}
\usepackage{amsmath}
\usepackage{amsthm}
\usepackage{mdwmath}
\usepackage{amsfonts}
\usepackage{graphicx}
\usepackage{graphics}
\usepackage{epstopdf}
\usepackage[mathscr]{euscript}
\usepackage{mdwmath}
\usepackage{bbm}
\usepackage{mathrsfs}
\usepackage{bm}
\usepackage[tight,footnotesize]{subfigure}
\usepackage{comment}
\usepackage{marginnote}
\usepackage{color}
\usepackage{tabularx}
\usepackage[flushleft]{threeparttable}
\usepackage{algorithm}
\usepackage[]{algpseudocode}
\usepackage{float}
\makeatletter
\newcommand{\algmargin}{\the\ALG@thistlm}
\makeatother
\newlength{\whilewidth}
\algdef{SE}[parWHILE]{parWhile}{EndparWhile}[1]
{\parbox[t]{\dimexpr\linewidth-\algmargin}{%
		\hangindent\whilewidth\strut\algorithmicwhile\ #1\ \algorithmicdo\strut}}{\algorithmicend\ \algorithmicwhile}%
\algnewcommand{\parState}[1]{\State%
	\parbox[t]{\dimexpr\linewidth-\algmargin}{\strut #1\strut}}
\algnewcommand{\parComment}[1]{\Comment%
	\parbox[t]{\dimexpr\linewidth-\algmargin}{\strut #1\strut}}
\theoremstyle{plain}

\newcommand{\Cc}{\mathbb{C}}

\begin{document}

\title{On Distribution Patterns of Power Flow Solutions}
\author{
\IEEEauthorblockN{Dan Wu, \IEEEmembership{Member, IEEE}, Bin Wang*, \IEEEmembership{Member, IEEE}, Kai Sun, \IEEEmembership{Senior Member, IEEE}, Le Xie, \IEEEmembership{Senior Member, IEEE}}
\thanks{D. Wu is with the Laboratory for Information and Decision Systems, Massachusetts Institute of Technology, Cambridge, MA, (email: danwumit@mit.edu).

B. Wang and L. Xie are with the Department of Electrical and Computer Engineering, Taxes A\&M University, College Station, TX, (emails: {binwang,le.xie}@tamu.edu

K. Sun is with the Department of Electrical Engineering and Computer Science, University of Tennessee, Knoxville, TN, (email: kaisun@utk.edu)}
}
\maketitle
\begin{abstract}
A fundamental challenge in computer analysis of power flow is the rigorous understanding of the impact of different loading levels on the solutions of the power flow equation. This letter presents a comprehensive study of possible numerical solutions that may arise as the loading level varies. In particular, a type of ``false" load flow solutions is reported for the first time as a legitimate numerical solutions but with no engineering justification. The existence and the mechanism of why this category of solutions exist are rigorously analyzed. The probability mass function of voltage solution is shown to be less dependent on loading levels. Furthermore, numbers of all actual solutions and those solutions within engineering limits are summarized. This letter presents a first attempt to compute such huge numbers of solutions to the tested systems, and analyze their distribution patterns. \emph{All solution sets investigated in this letter are posted online associated with this letter to support any further research purposes.} 
\end{abstract}


\section{Introduction}
\label{sec:intro}
The increasing integration of distributed energy resources and demand responses provides a great flexibility to operate power grids in a more efficient manner. However, this flexibility can also alter the traditional load pattern and pattern diversity. This letter provides a comprehensive study to better understand the influence on the power flow solutions. 

It is well-known that power balance equations can admit multiple solutions. Other than the high-voltage solution, most of the rest do not allow a stable and secure operation. However, their locations and distributions convey important information about the underlying system, shown to be useful for both static and dynamical stability analysis \cite{tamura1983pfsoluvsa,chiang2011:direct}. Finding multiple power flow solutions is not a trivial task. It was computationally tractable only up to systems with 14 buses \cite{mehta2016:numerical}. A recently developed technique, called \emph{holomorphic embedding based continuation} (HEBC) method \cite{wu2019:hebc}, pushes this boundary to a system with 57 buses. Currently, there is still a lack of theoretical proof to show if the HEBC method can find all power flow solutions. However, HEBC method always gives complete solution sets in all verifiable cases \cite{wu2019:hebc}. 

This letter adopts the HEBC method to find multiple power flow solutions. 
To assess potential statistical properties of power flow solutions, a few cases, i.e. Case14, Case30, Case39 and Case57, are adopted in this letter leading to a comprehensive investigation. Their solution sets are posted online with this letter\footnote{Dataset DOI provided by IEEE DataPort: 10.21227/24bh-hj72. This dataset can also be accessed at https://danwu.mit.edu/power-flow.}.

\section{``Short Circuit'' Solutions}
\label{sec:sc}
A specific type of power flow solutions is observed. These particular solutions satisfy the power balance equations, namely the power injection model, and thus are the solutions to the power flow problem. But they do not satisfy the Kirchhoff's Current Law (KCL) as explained in the following. We refer to them as the \emph{``short circuit'' solutions}. A necessary condition for the existence of ``short circuit'' solution is that: \emph{a power grid has at least one transit-bus which is a special PQ-bus with zero nodal power injection.} 

Consider the power balance equation at node-$i$, 
\begin{equation}
	V_i \times I_i^\star = P_i + jQ_i \label{eq:power}%
\end{equation}%
where $V_i \in \Cc$ is the complex voltage at node-$i$, $I_i^\star \in \Cc$ is the conjugate of complex current at node-$i$, $P_i + jQ_i  \in \Cc$ is the complex power injection at node-$i$.

If node-$i$ is a transit-bus, then $P_i + jQ_i  = 0$, which suggests that either $V_i=0$ or $I_i=0$\footnote{It is possible that both $V_i$ and $I_i$ vanish to zero. However, this is rather rare and hasn't been observed in our studies.}. When $V_i=0$ and $I_i\neq 0$, an external current $I_i$ is supposed to enter node-$i$ from somewhere. Since $V_i=0$ at this point, the solution suggests that the system is grounded at node-$i$. However, the system is not physically connected to the ground at node-$i$. Therefore, KCL fails at this solution. One can interpret this ``short circuit'' solution as a feasible solution to the grounded power flow problem at node-$i$. Reference \cite{milano2010:scripting} reported a numerical problem caused by ``short circuit'' solutions, i.e. voltage can be wrongly trapped at zero after clearing a short-circuit fault, when using power injection model in dynamic simulations.

\begin{figure}[!ht]
	\centering
	\subfigure[Case30]{\label{fig:case30scs}\includegraphics[width=0.47\columnwidth]{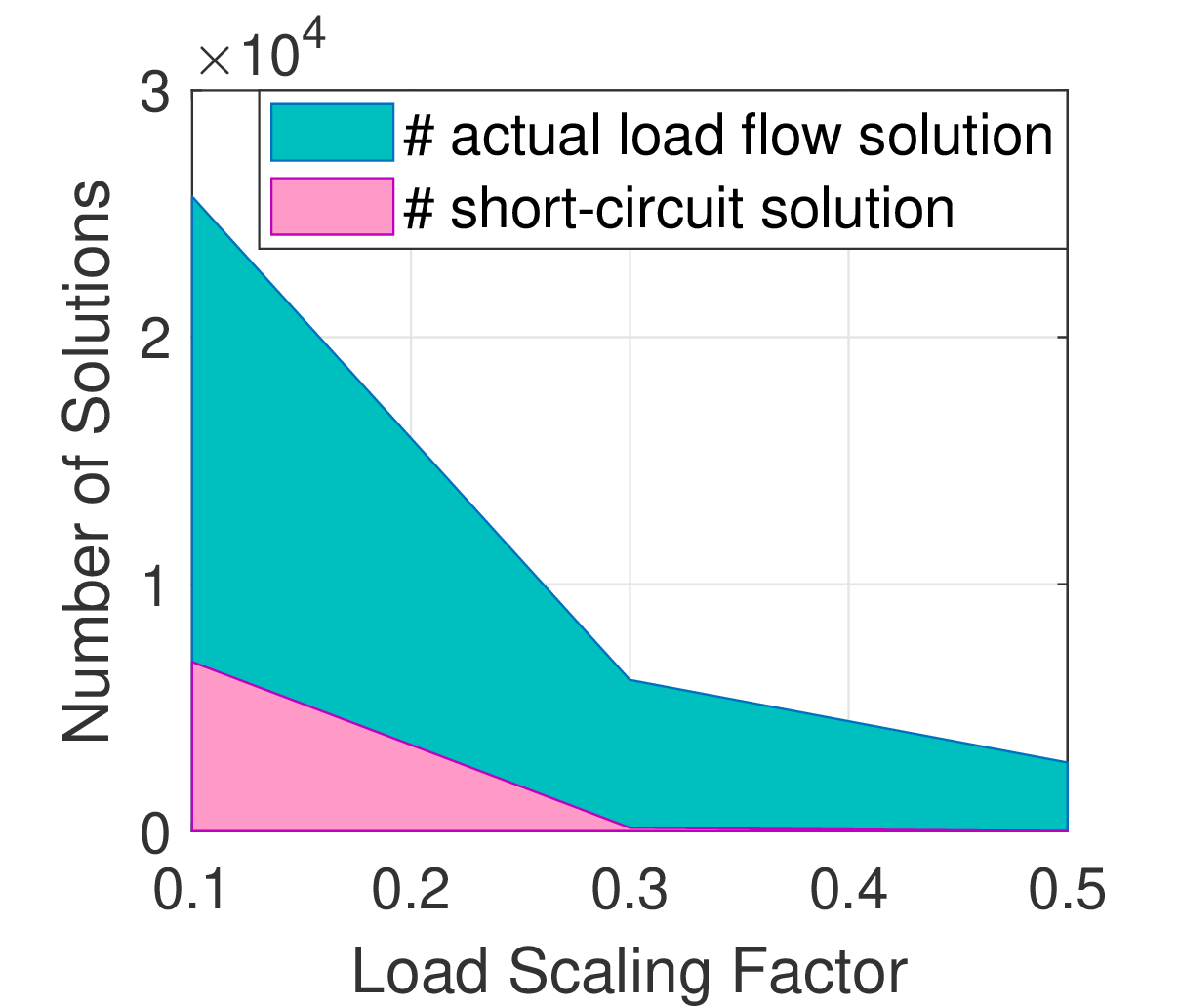}}~
	\subfigure[Case39 ]{\label{fig:case39scs}\includegraphics[width=0.47\columnwidth]{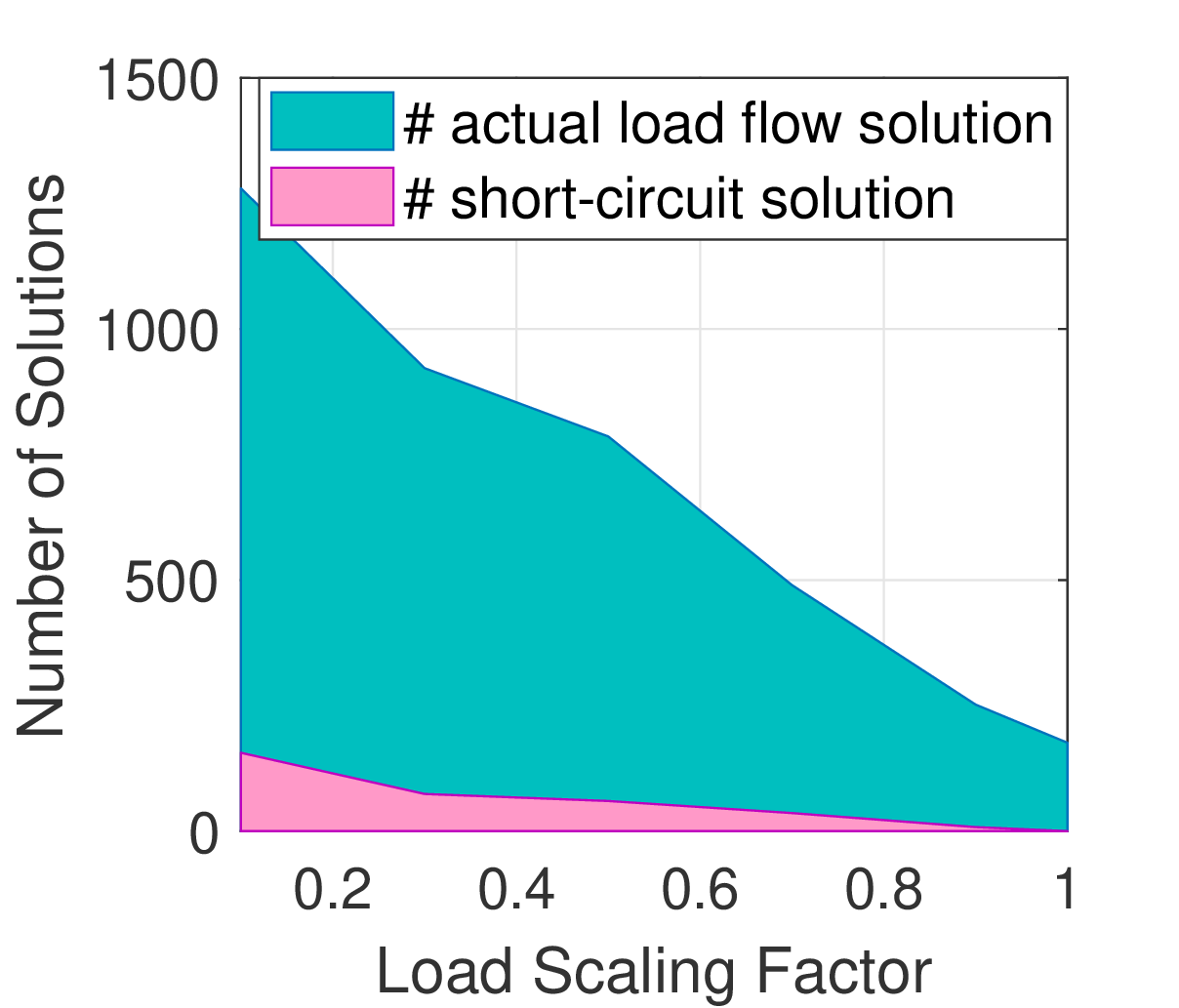}}
	\caption{Number of Short-Circuit Solutions} \label{fig:scs}%
\end{figure}

Our studies show that ``short circuit'' solutions are more likely to happen at a lighter loading level. For example, Fig.~\ref{fig:scs} shows the number of ``short circuit'' solutions with respect to the load scaling factor\footnote{The load scaling factor scales every complex-valued load in the system.}. When the load increases to a certain level, they eventually disappear. But at low loading levels, the number of ``short circuit'' solutions can be huge. For instance, in Fig. \ref{fig:case30scs}, Case30 has at least $6849$ ``short circuit'' solutions at $10 \%$ loading level. A light loading level usually admits much more ``short circuit'' solutions because the constant loading line can intersect with many PV (QV) branches for the grounded power flow problem.

``Short circuit'' solutions can also occur when constant impedance loads exist. These load nodes are transit-bus, and thus can admit many such solutions at a light loading level. 

A simple way to avoid ``short circuit'' solutions is to add a small power injection, say, $10^{-5} p.u.$\footnote{In our simulations the power mismatch error is below $10^{-9}$ p.u. Hence a $10^{-5}$ power injection will not be confused by the error threshold.}, at each transit-bus. As long as the power injection at each node-$i$ in \eqref{eq:power} is non-zero, neither $V_i$ nor $I_i$ is zero. 

\section{Numerical Study of Solution Distribution}
\label{sec:num}

This section first illustrates that the number of power flow solutions decreases with the increase of loading level. It is then shown that under lightly-loaded conditions, the identified solutions in each case are not randomly distributed, but exhibit distinct patterns. A scrutiny on the identified solution sets in terms of voltage magnitudes of PQ buses and reactive power of PV buses is presented.

\subsection{Number of Solutions at Different Loading Levels}

\begin{figure}[!ht]
	\centering
	\subfigure[Case14]{\label{fig:14base}\includegraphics[width=0.45\columnwidth]{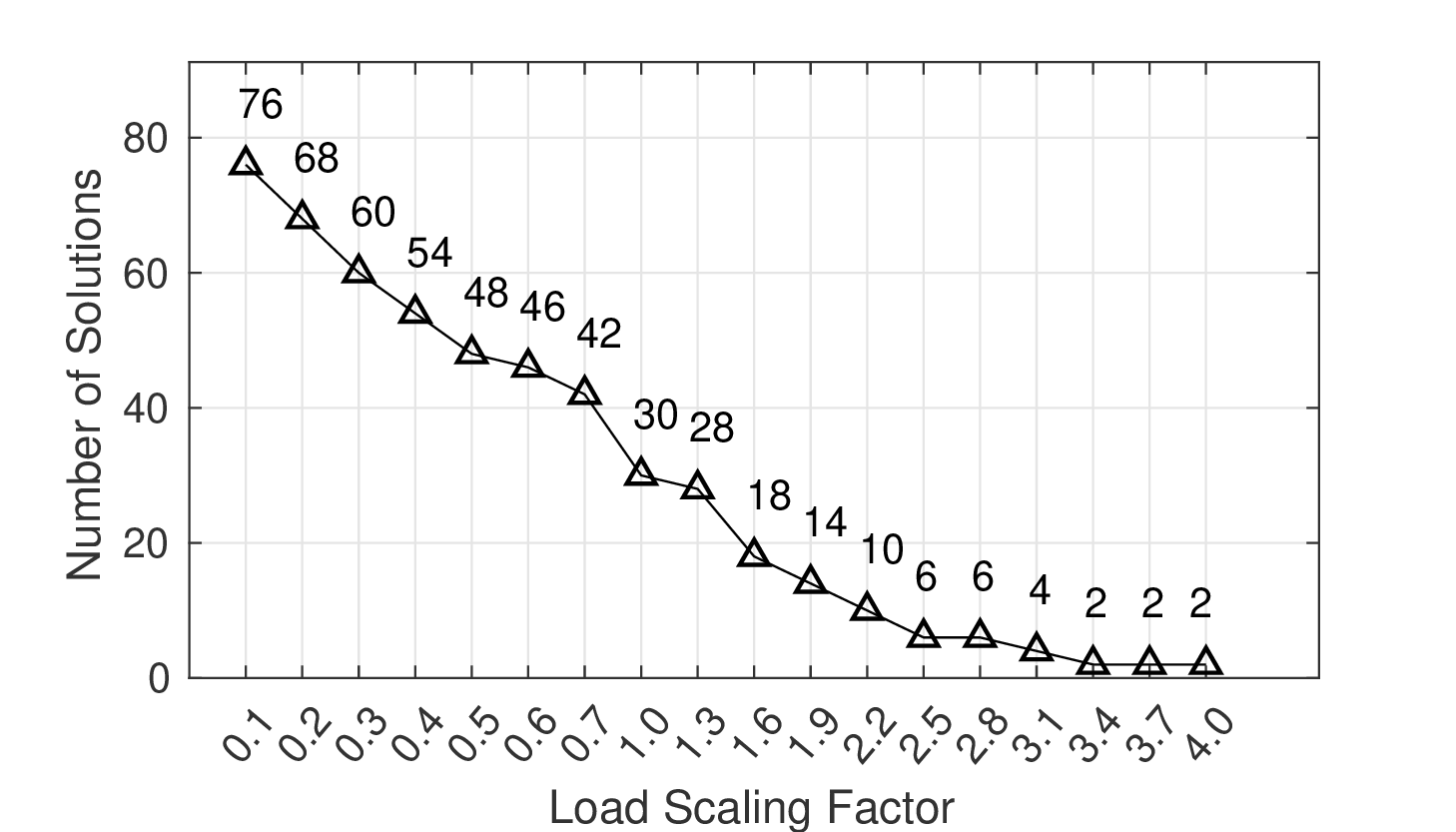}}~~~~
	\subfigure[Case30 ]{\label{fig:30base}\includegraphics[width=0.45\columnwidth]{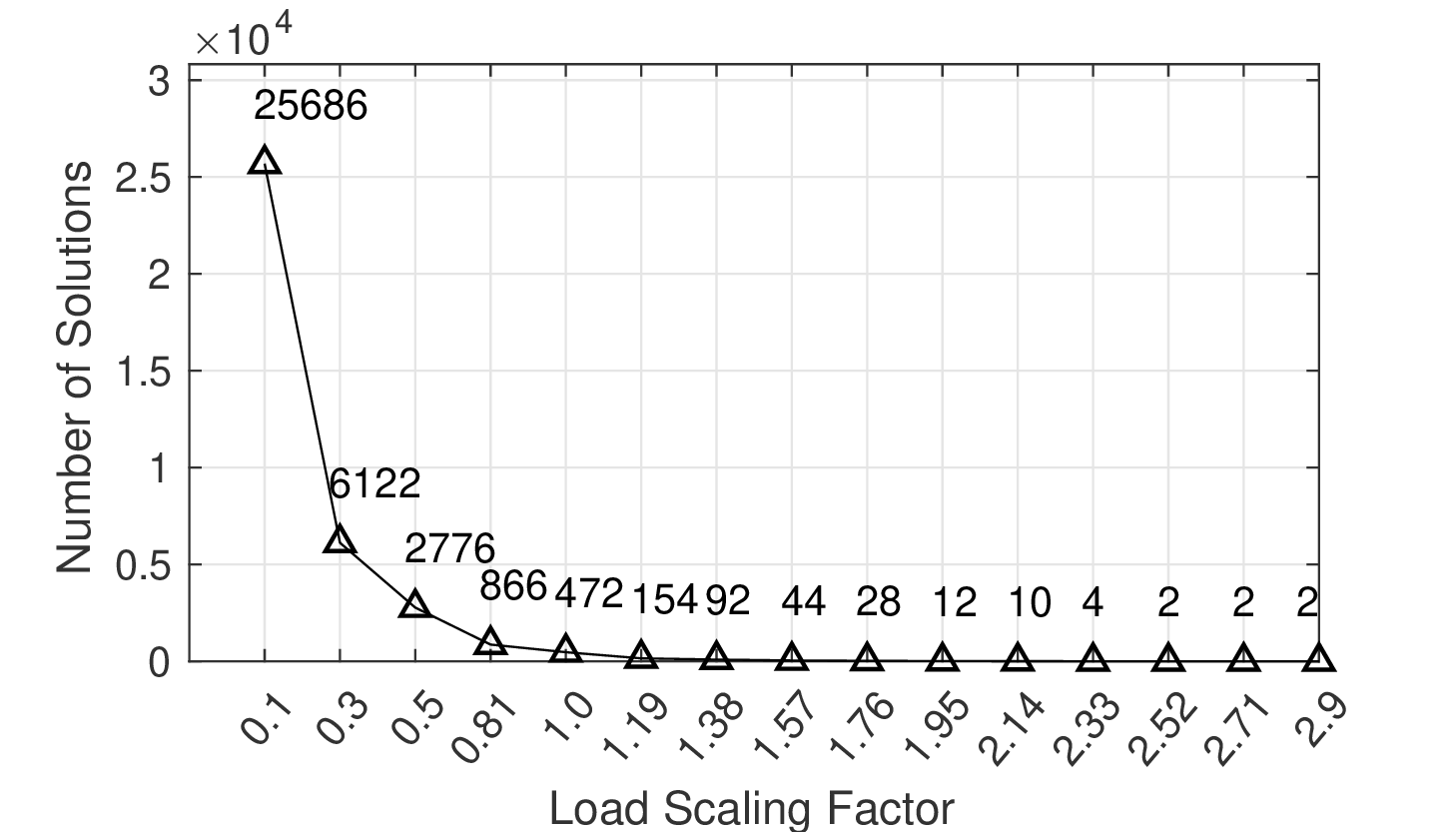}}\\
	\subfigure[Case39]{\label{fig:39base}\includegraphics[width=0.45\columnwidth]{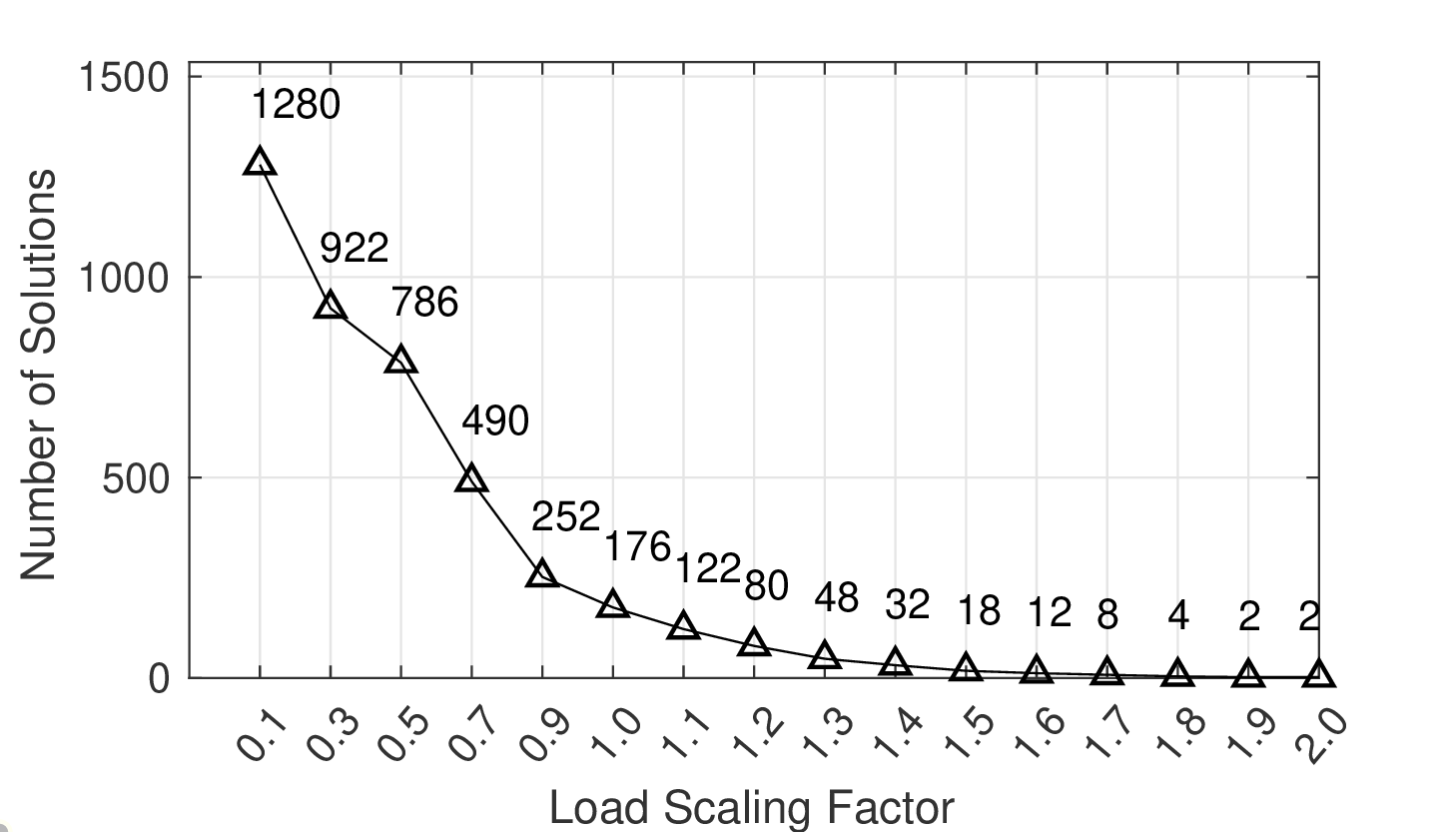}}
	~~~
	\subfigure[Case57]{\label{fig:57base}\includegraphics[width=0.45\columnwidth]{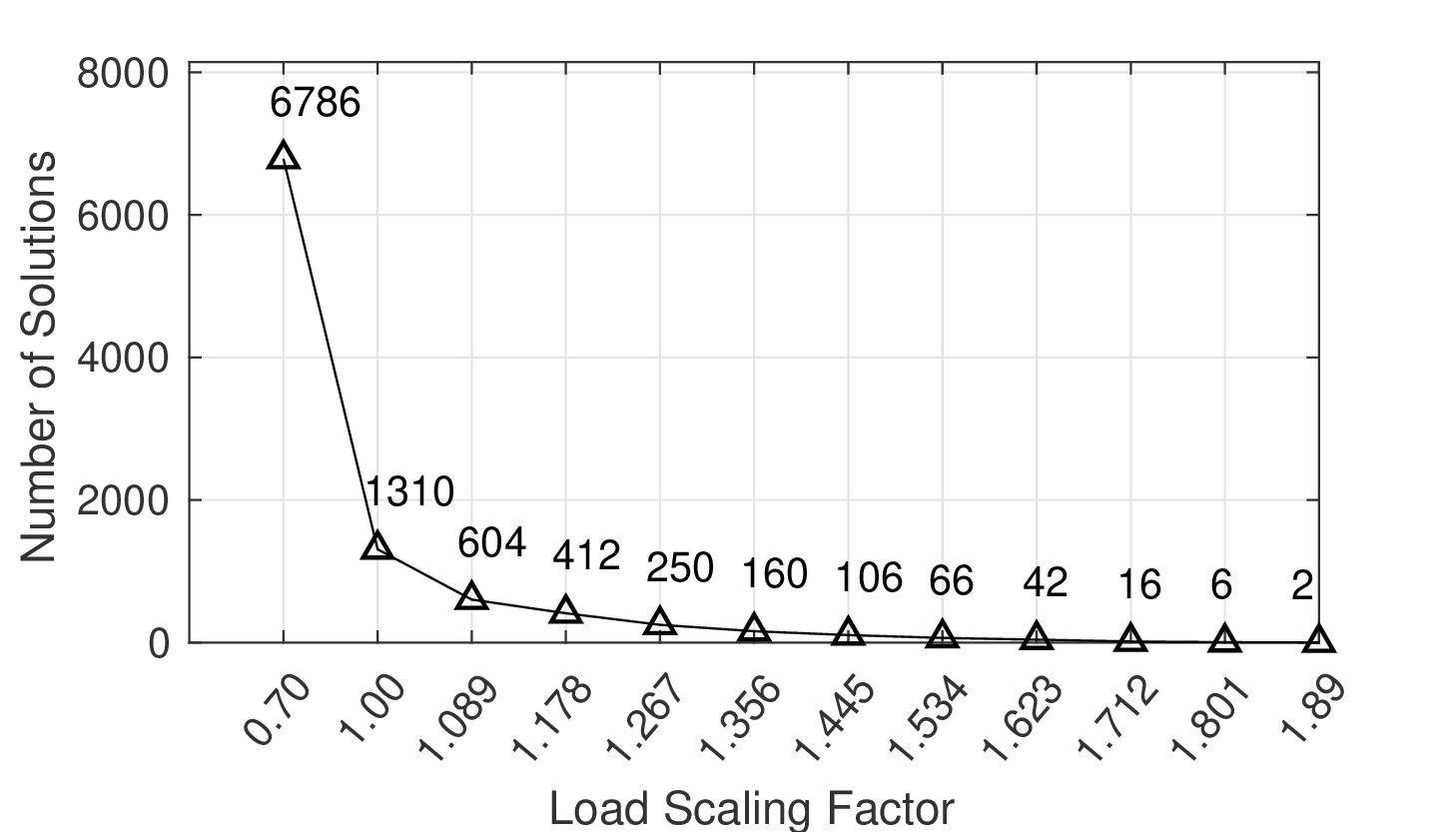}}
	\caption{Number of power flow solutions as loading level increases} \label{fig:loadscale}
\end{figure}

It is expected that as the loading level increases, the number of power flow solutions decreases, e.g. down to two solutions right before the voltage collapse at the saddle node as illustrated in Fig. \ref{fig:loadscale}. On the other hand, at light loading levels, the number of solutions can be very huge. For instance, Fig. \ref{fig:30base} shows that $10\%$ loading induces at least $25686$ power flow solutions for Case30. This huge number, probably even greater for lighter loading or larger cases, makes it very challenging and less attractive to find all associated power flow solutions because the system stability is usually of less a concern under light loading conditions. At extremely heavy loading conditions, however, the number of solutions can be very small. Therefore, it is more practical and beneficial to identify these solutions for stressed power systems.

\subsection{Node Voltage Pattern}
\begin{figure}[!ht]
	\centering
	\subfigure[Pattern 1 ]{\label{fig:case30010v16}\includegraphics[width=0.45\columnwidth]{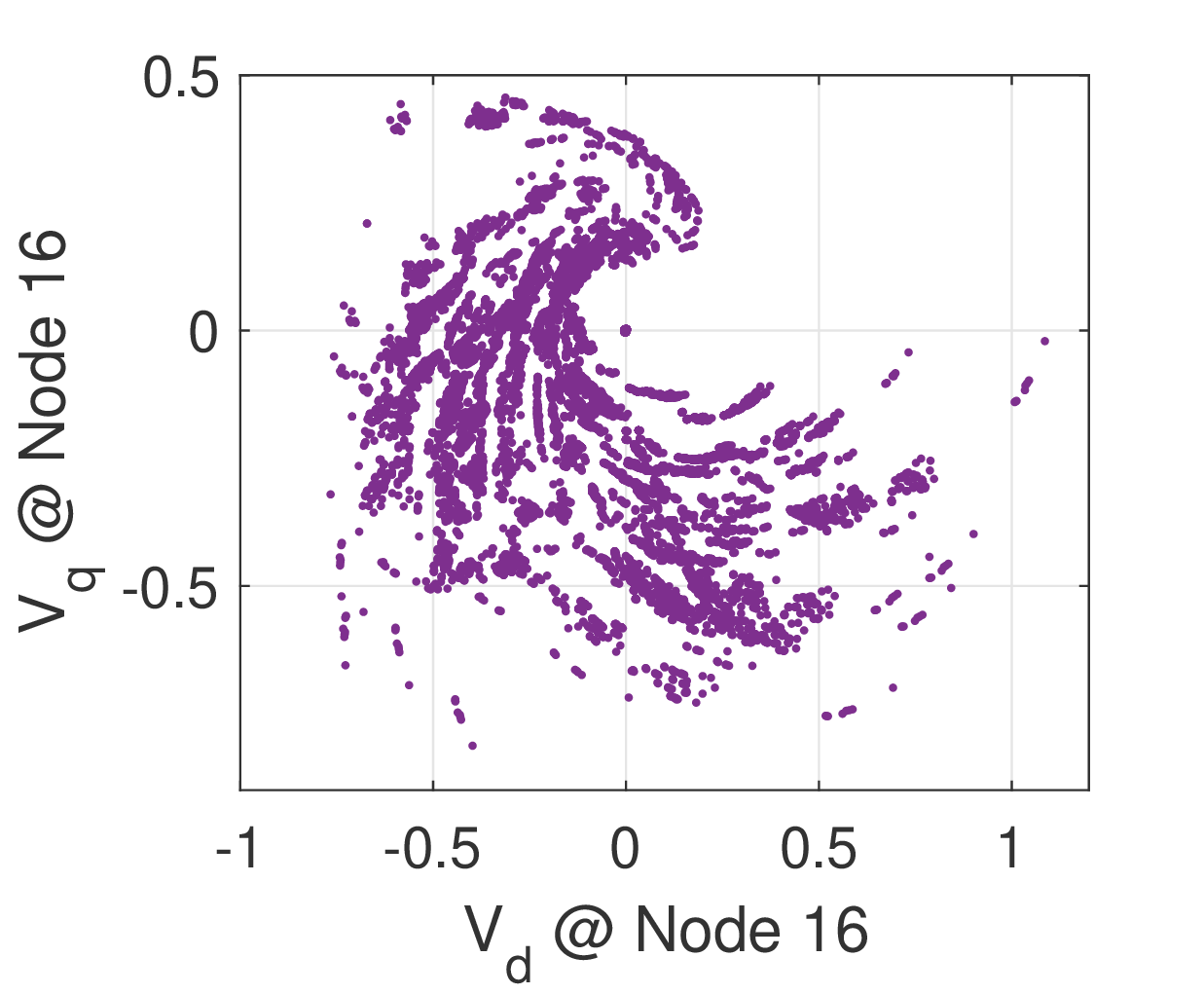}}~
	\subfigure[Pattern 2 ]{\label{fig:case30010v21}\includegraphics[width=0.45\columnwidth]{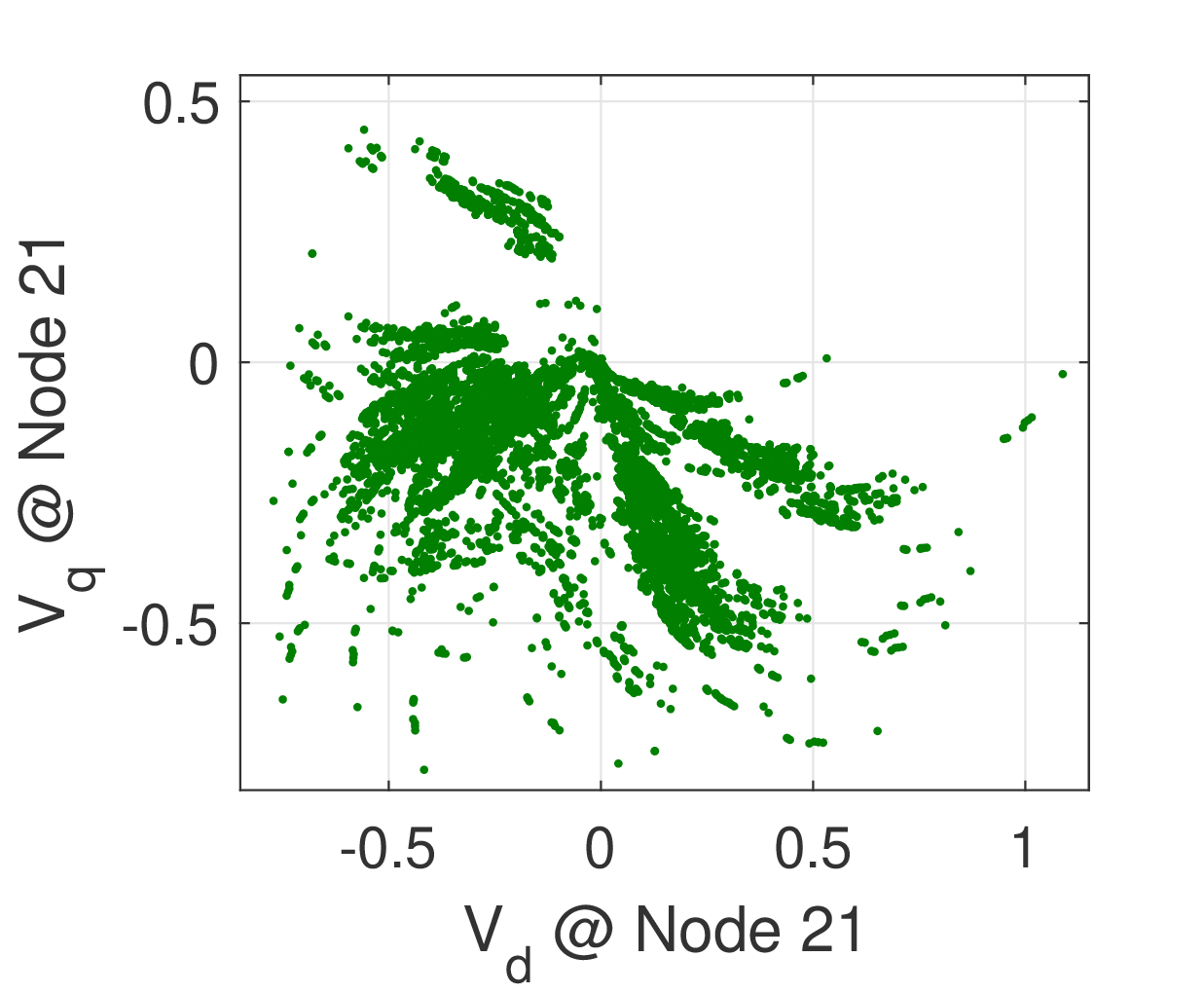}}\\
	\subfigure[Pattern 3]{\label{fig:case30010v29}\includegraphics[width=0.45\columnwidth]{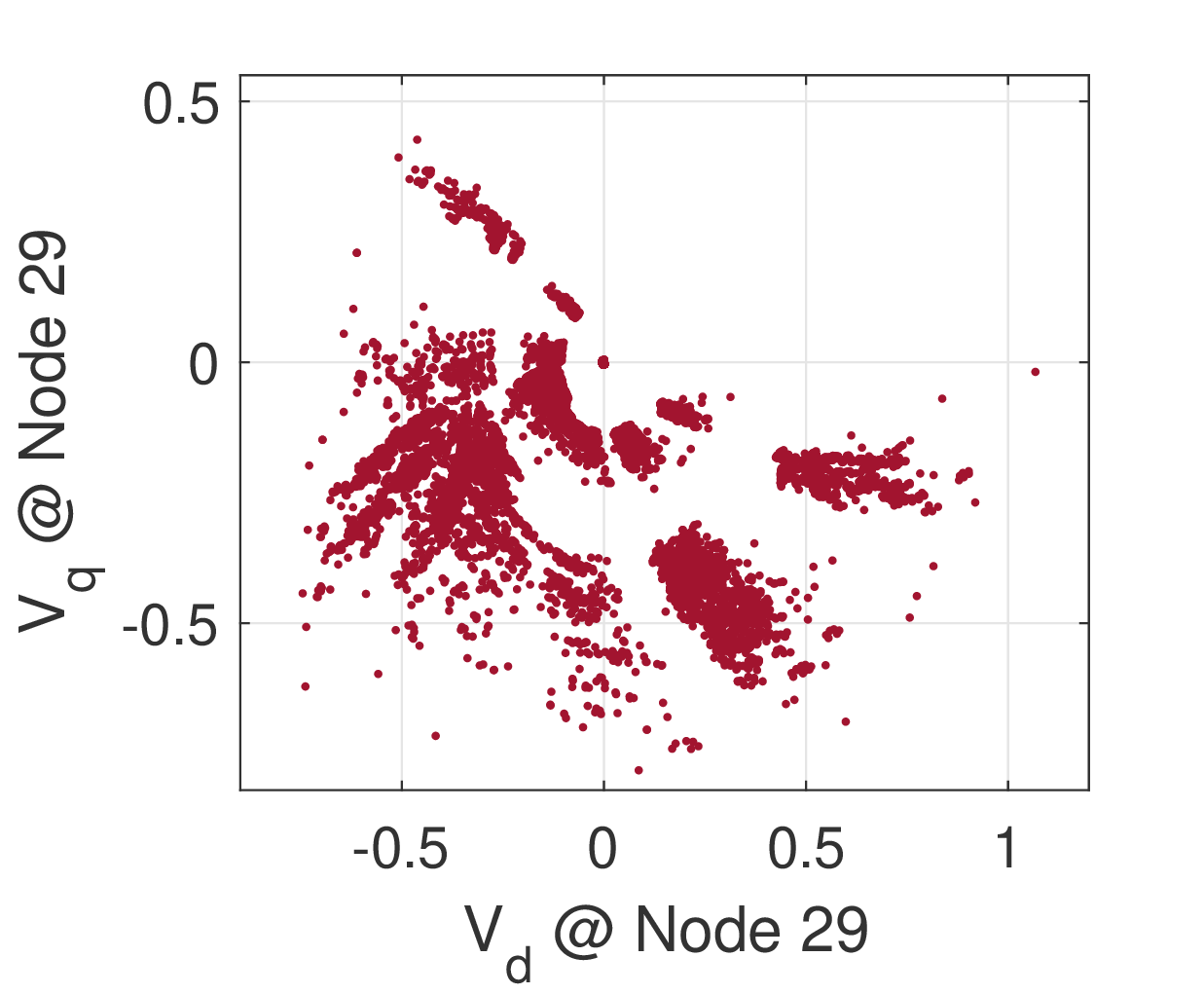}}~
	\subfigure[Pattern 4 ]{\label{fig:case30010v4}\includegraphics[width=0.45\columnwidth]{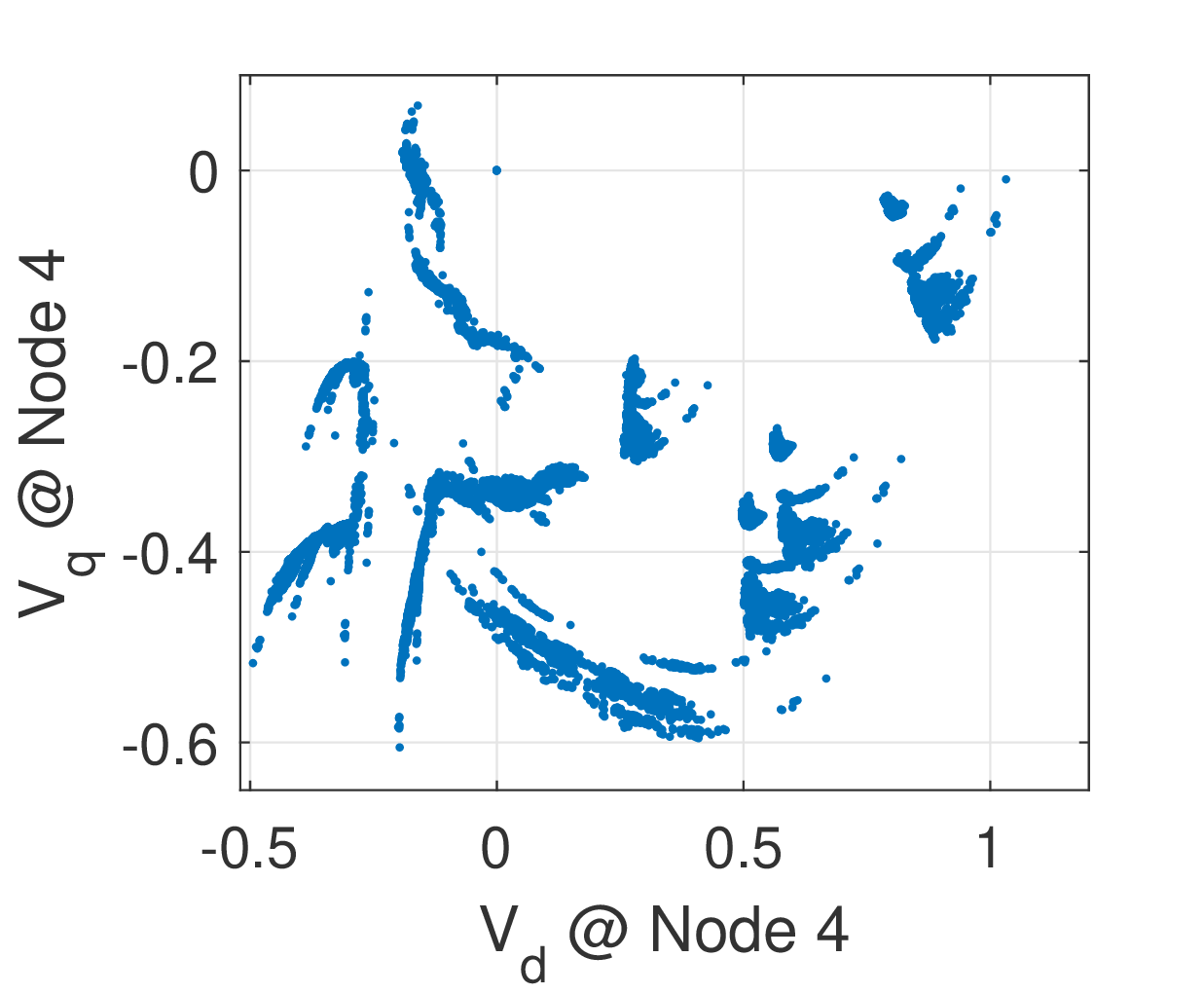}}
	\caption{Voltage Patterns for Case30 at $10\%$ Load (25686 Solutions)} \label{fig:case30}
\end{figure}

An interesting observation is that for each test case the nodes can be clustered by a few special voltage patterns. For example, Fig. \ref{fig:case30} depicts four basic patterns that occur in Case30, where each dot represents a voltage solution. The horizontal and vertical axes respectively represent the real and imaginary parts of the complex voltage. Table~\ref{table:1} summarizes the clusters of nodes that exhibit similar patterns as shown in Fig. \ref{fig:case30}. These structures persist as the loading condition changes. A light loading level, i.e. $10\%$, is adopted here since it gives a sufficient number of power flow solutions for exhibiting potential statistical properties. A common pattern omitted here is a fixed-radius circle which is associated with each PV bus.

\begin{figure}[!ht]
	\centering
	\subfigure[Case39 $30 \%$ Load ]{\label{fig:case39030v2}\includegraphics[width=0.48\columnwidth]{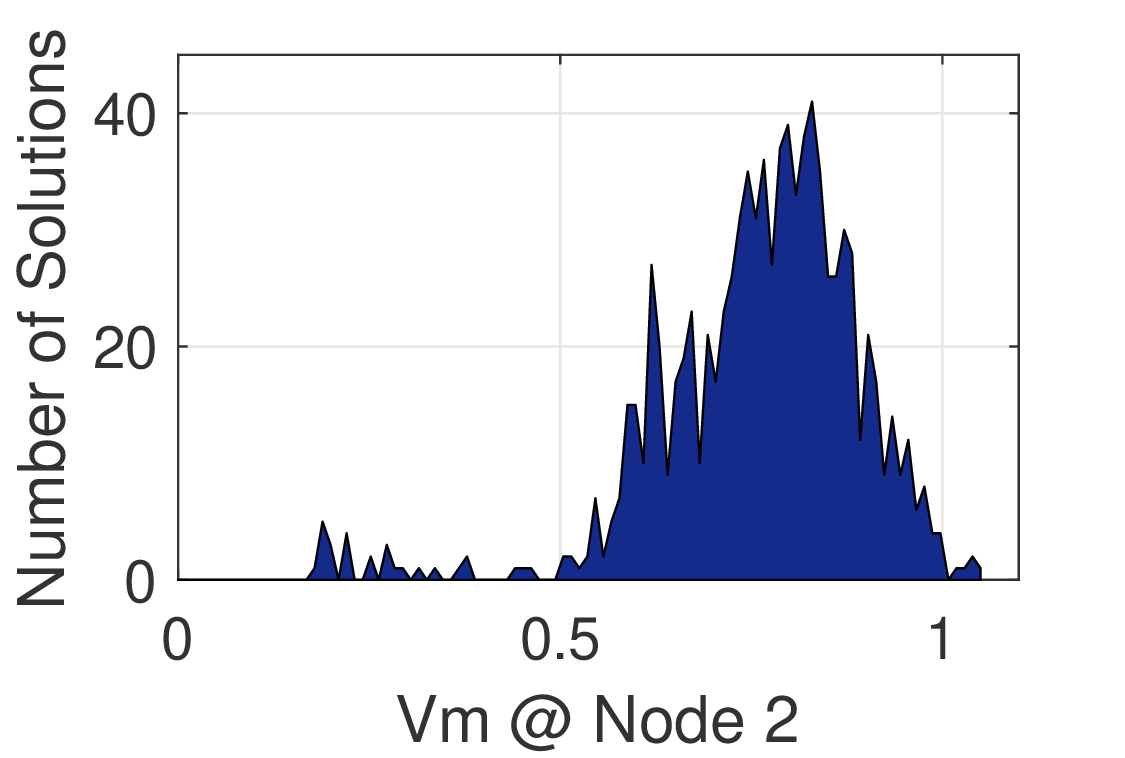}}~
	\subfigure[Case39 $90 \%$ Load ]{\label{fig:case39090v2}\includegraphics[width=0.48\columnwidth]{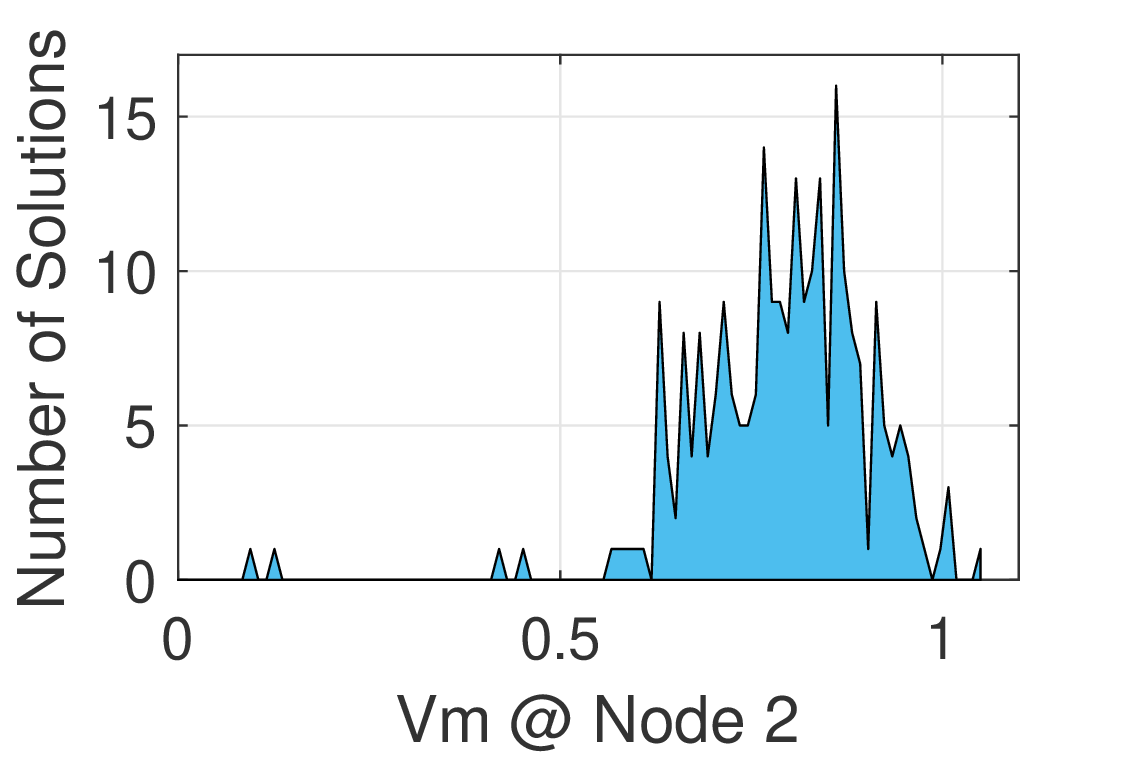}}\\
	\subfigure[Case57 $70 \%$ Load ]{\label{fig:case570700v56}\includegraphics[width=0.48\columnwidth]{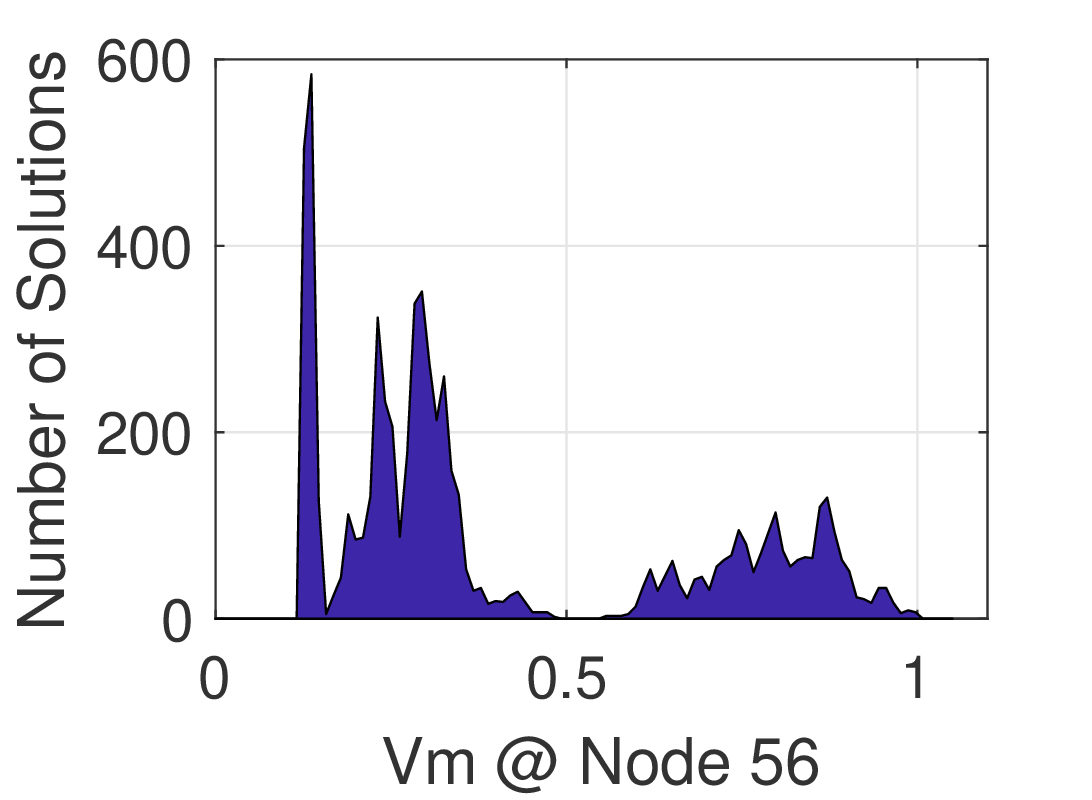}}~
	\subfigure[Case57 $100 \%$ Load ]{\label{fig:case571000v56}\includegraphics[width=0.48\columnwidth]{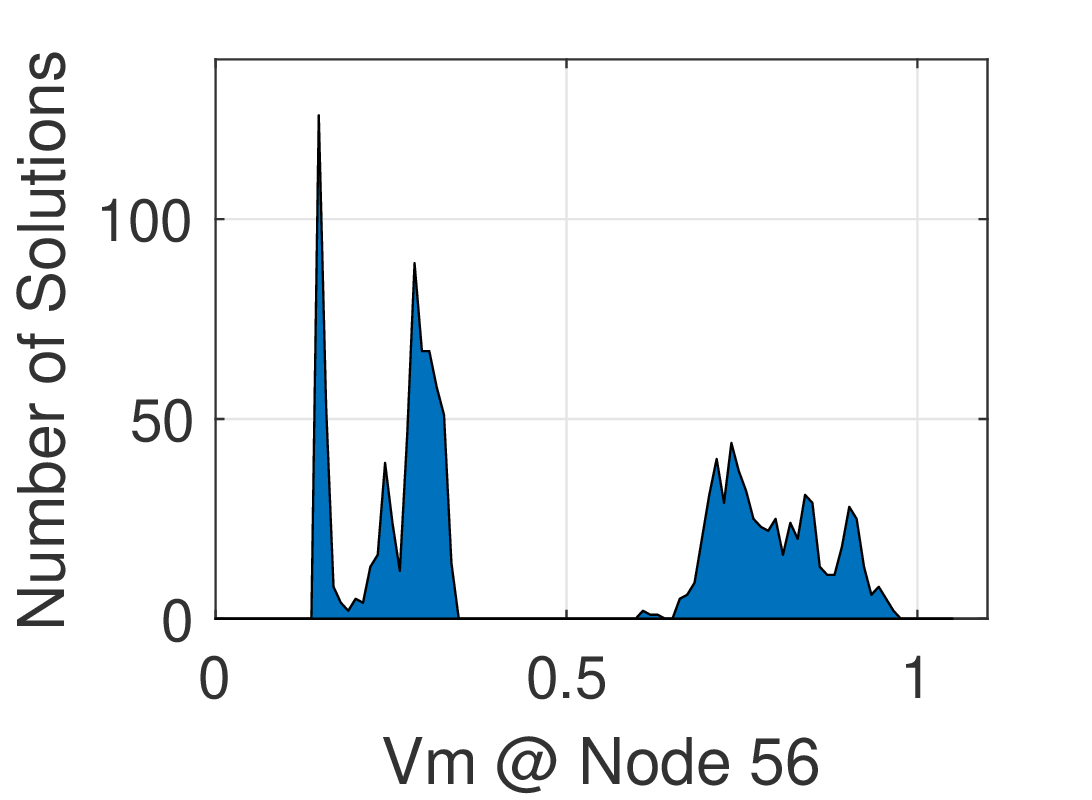}}
	\caption{Voltage Magnitude Distributions} \label{fig:case39_distr}
\end{figure}
\begin{table}[!ht]
	\caption{Bus Grouping by Voltage Pattern of Case30} \label{table:1}
	\begin{tabular}{c|c|c|c|c}
		\hline
		Patterns & 1         & 2     & 3                & 4          \\ \hline
		Buses    & 12, 14-16 & 17-24 & 9,10,25-27,29,30 & 3,4,6,7,28 \\ \hline
	\end{tabular}
\end{table}

The 39-bus system and the 57-bus system also exhibit a few distinctive voltage patterns. Usually, adjacent nodes are more likely to share the same pattern, but it is not always the case. Whether different patterns reveal local structural properties of the system is an open question. But the persistence of these patterns under different loading conditions may suggest a relation with the network topology. Some of the patterns are further depicted in Subsection-D with engineering limit considerations. 

To reveal the statistical characteristics of power flow solutions, we discretized the voltage magnitude range $[0,1.1]~p.u.$ for $100$ even intervals, and count the number of solutions for each interval. Sample distributions are depicted in Fig. \ref{fig:case39_distr}, which exhibits persisting patters that are consistent with the observations in fig.~\ref{fig:case30}.


\subsection{Solutions within Engineering Limits}
Although there exist many power flow solutions, it turns out that only very few of them can comply with engineering limits, including bus voltage limit and generator reactive power limit. 

\begin{figure}[!ht]
	\centering
	\subfigure[Node 2 of Case39 at $10\%$ Load ]{\label{fig:case39s1}\includegraphics[width=0.48\columnwidth]{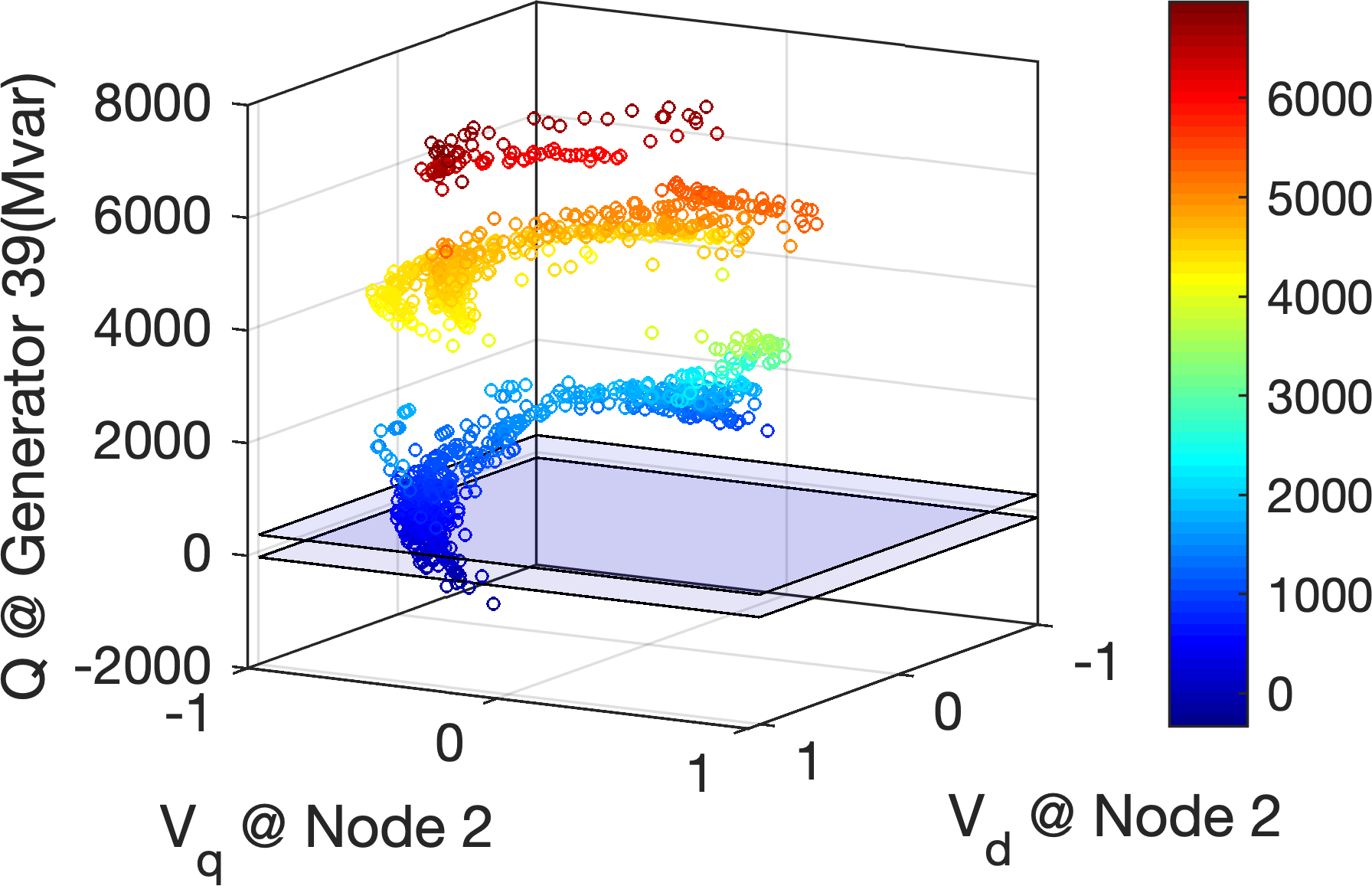}}~
	\subfigure[Node 28 of Case39 at $10\%$ Load ]{\label{fig:case39s2}\includegraphics[width=0.48\columnwidth]{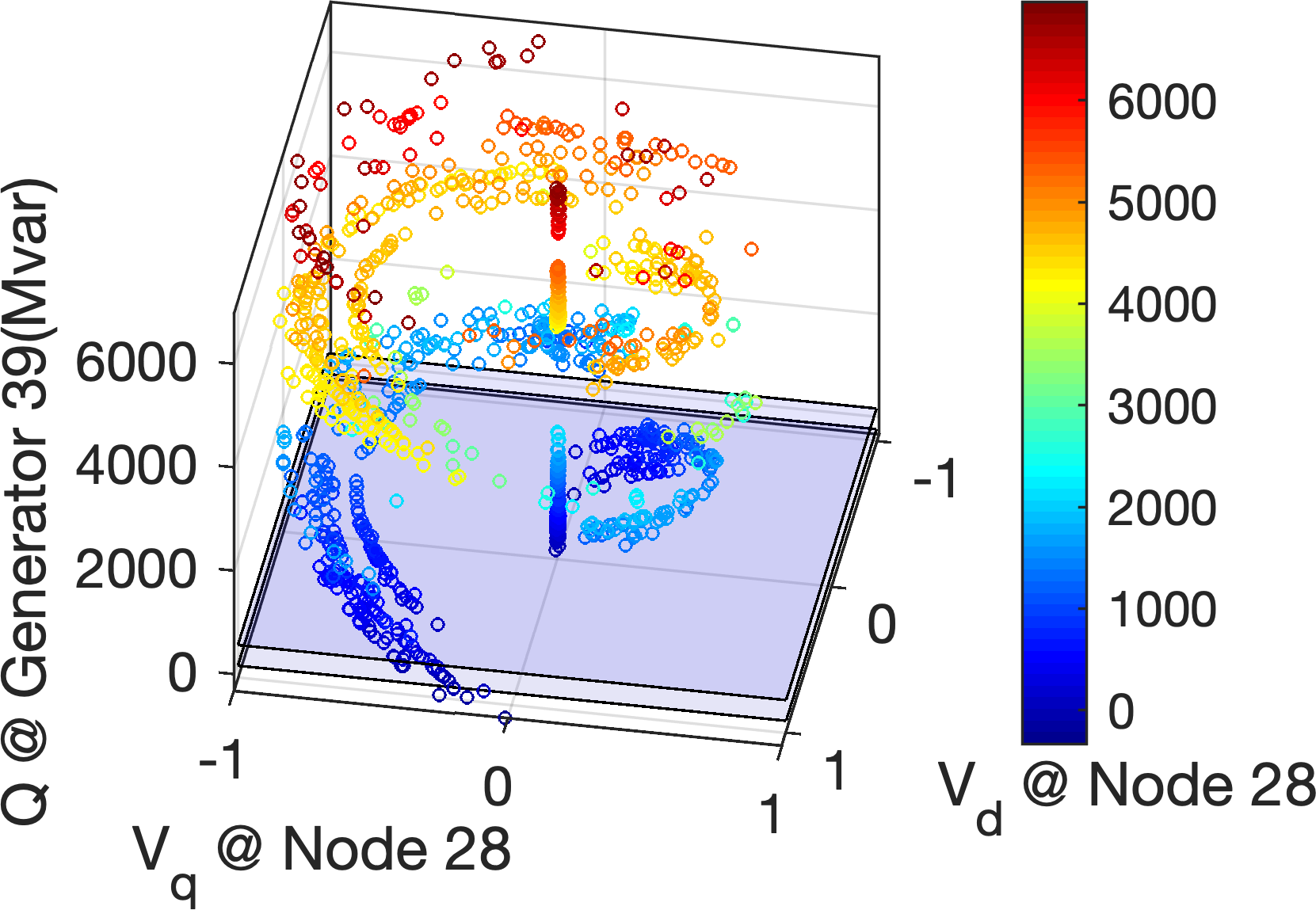}}\\
	\subfigure[Node 31 of Case57 at $70\%$ Load]{\label{fig:case57s1}\includegraphics[width=0.48\columnwidth]{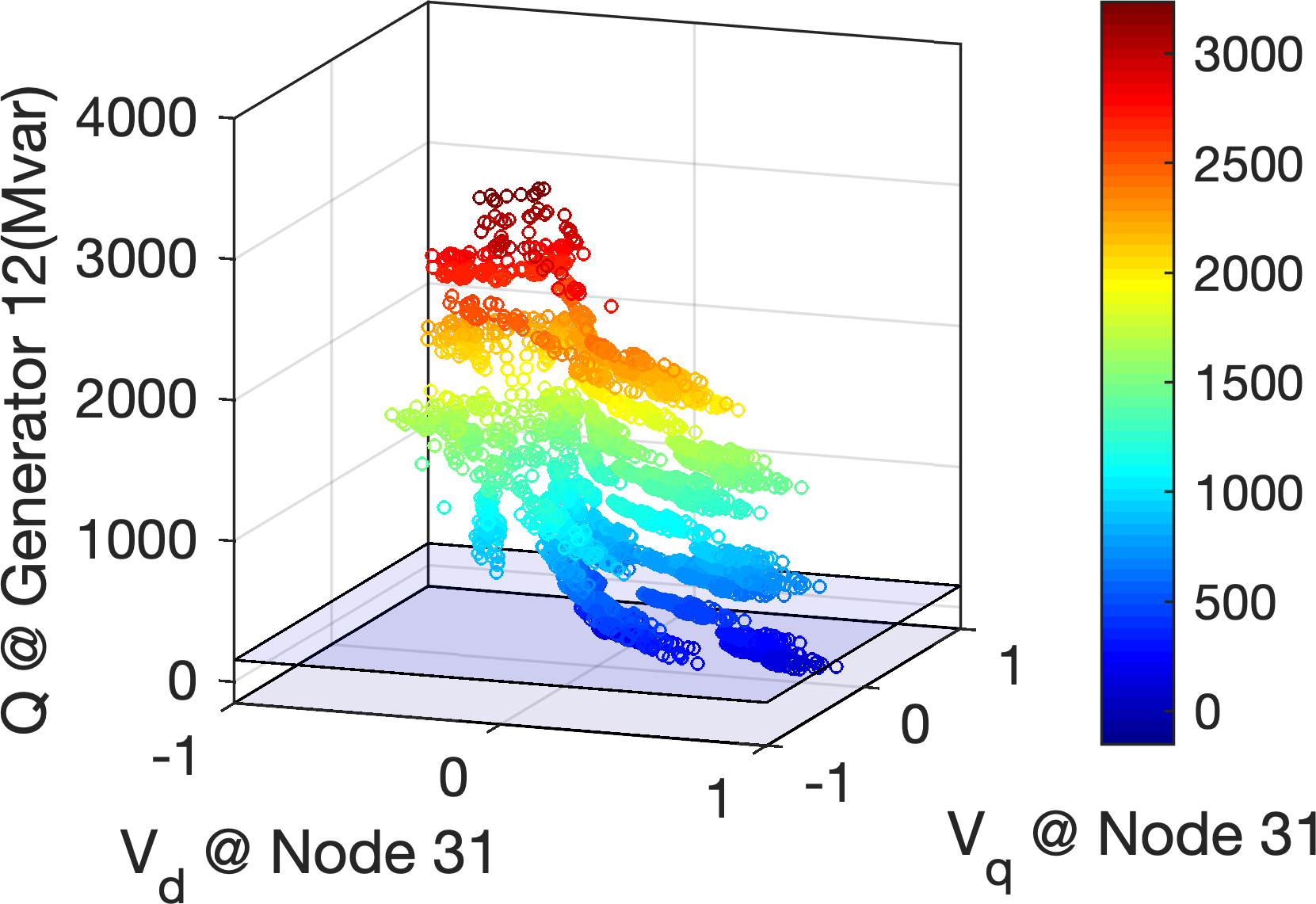}}~
	\subfigure[Node 55 of Case57 at $70\%$ Load]{\label{fig:case57s2}\includegraphics[width=0.48\columnwidth]{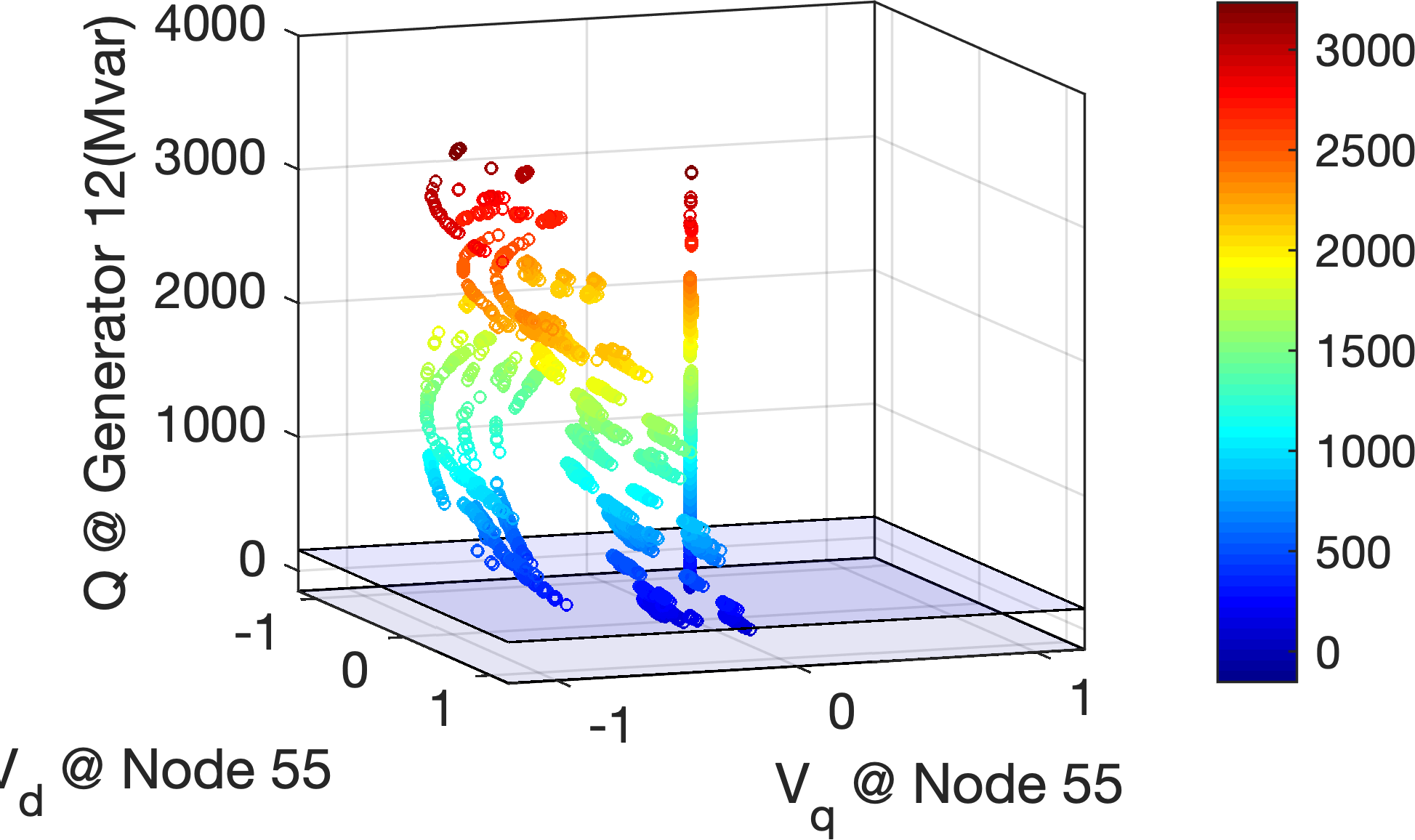}}\\
	
	\caption{Bus Voltage v.s. Generator Reactive Power} \label{fig:case3957}
\end{figure}

Fig. \ref{fig:case3957} extends Fig. \ref{fig:case30} by adding the reactive power output of a certain generator as the z-axis. The two planes parallel with the plane of $X-Y$, representing practical reactive power limits, cut each 3-D space into three regions, where the region in the middle is a secure region with no reactive power violation. In Case39 with $10\%$ load, there is only 1 solution (out of $1280$ solutions) complying with reactive power limit at each generator. In Case57 with $70\%$ load, only $23$ out of the total $6786$ solutions have their reactive power outputs on PV buses within limits. In addition, another interesting phenomenon can be observed in the 3-D plots in Fig. \ref{fig:case3957} that power flow solutions are basically distributed on a spiral surface stretching along z-axis.   

\begin{figure}[!ht]
    \centering
    \subfigure[Case39 at $10\%$ Load]{\label{fig:case39_vmmin}\includegraphics[width=0.48\columnwidth]{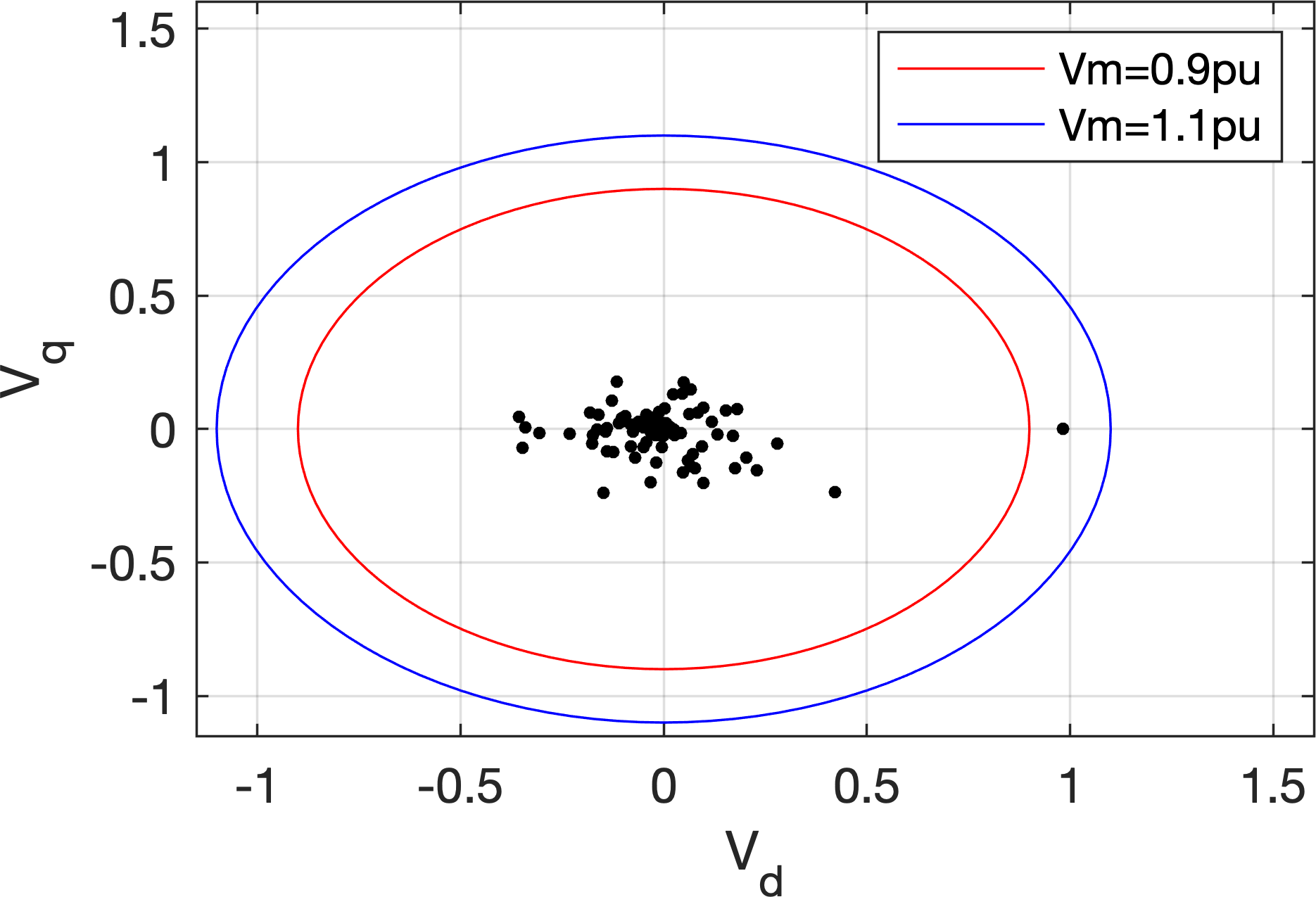}}~
    \subfigure[Case57 at $70\%$ Load]{\label{fig:case57_vmmin}\includegraphics[width=0.48\columnwidth]{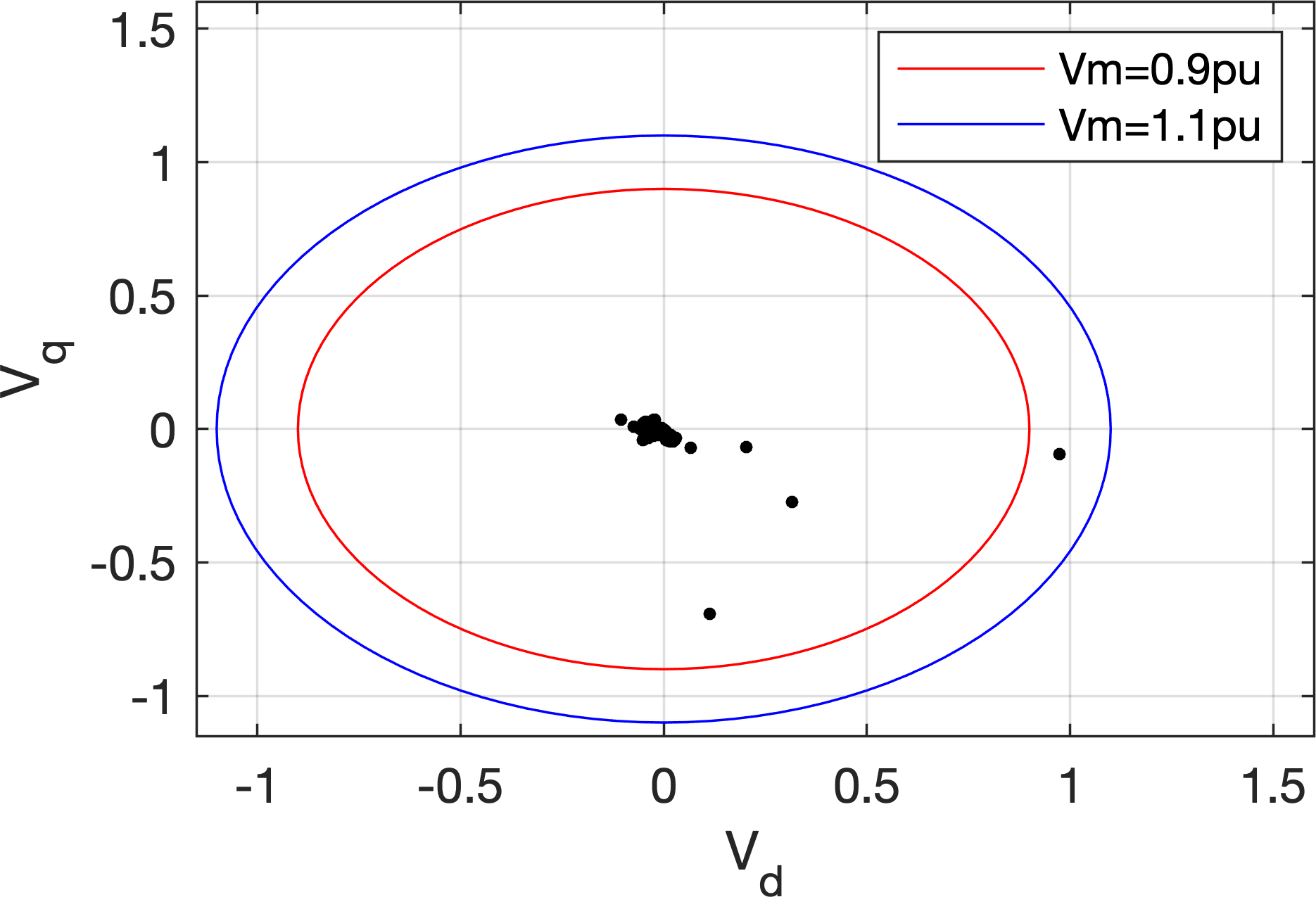}}
    \caption{Minimum Bus Voltage Magnitude of Different Solutions}
    \label{fig:case3957minV}
\end{figure}

Fig. \ref{fig:case3957minV} shows that although a light loading gives many power flow solutions, there is usually only one solution with its minimum voltage magnitude higher than $0.9~p.u.$. It is worth mentioning that multiple high-voltage power flow solutions may exist under some special conditions or network structures \cite{nguyen}.

\section{Conclusion}
\label{sec:con}
The letter shows the existence of ``short circuit'' solutions to the power flow problem and analyzes distribution patterns of actual power flow solutions. All investigated solution sets are posted online with this letter. The presented results can serve as a benchmark for any further research purposes, including (i) investigation on the geometric structure of power balance equations; (ii) advanced computational methods for finding multiple power flow solutions; (iii) applications on static voltage stability and transient stability analyses using multiple power flow solutions.

\bibliographystyle{ieeetr}
\bibliography{references}

\begin{thebibliography}{1}

\bibitem{tamura1983pfsoluvsa}
Y.~Tamura, Y.~Nakanishi, and S.~Iwamoto, ``Relationship between voltage
  instability and multiple load flow solutions in electric power systems,''
  {\em IEEE Transactions on Power Apparatus and Systems}, vol.~PAS-102,
  pp.~1115--1123, 1983.

\bibitem{chiang2011:direct}
H.-D. Chiang, {\em Direct methods for stability analysis of electric power
  systems: theoretical foundation, BCU methodologies, and applications}.
\newblock John Wiley \& Sons, 2011.

\bibitem{mehta2016:numerical}
D.~Mehta, H.~D. Nguyen, and K.~Turitsyn, ``Numerical polynomial homotopy
  continuation method to locate all the power flow solutions,'' {\em IET
  Generation, Transmission \& Distribution}, vol.~10, no.~12, pp.~2972--2980,
  2016.

\bibitem{wu2019:hebc}
D.~Wu and B.~Wang, ``A holomorphic embedding based continuationmethod for
  identifying multiple power flowsolutions,'' {\em IEEE Access}, accepted,
  2019.

\bibitem{milano2010:scripting}
F.~Milano, {\em Power System Modelling and Scripting}.
\newblock Springer, Berlin, Heidelberg, 2010.

\bibitem{nguyen}
H.~D. {Nguyen} and K.~S. {Turitsyn}, ``Appearance of multiple stable load flow
  solutions under power flow reversal conditions,'' in {\em 2014 IEEE PES
  General Meeting | Conference Exposition}, pp.~1--5, July 2014.

\end{thebibliography}

\end{document}